\documentclass[12pt]{article}
\usepackage{amsmath,amssymb,amsbsy,amsfonts,amsthm,latexsym,amsopn,amstext,
            amsxtra,euscript,amscd}
\usepackage[latin1]{inputenc} 
\usepackage{algorithm}
\usepackage{algorithmic}
\usepackage{listings}
\usepackage{hyperref}

\begin{document}

\baselineskip 20pt

\begin{center} {\bf A simple coding-decoding algorithm for the Hamming code}
\end{center}

\begin{center} {Omar Khadir}

Laboratory of Mathematics, Cryptography, Mechanics and Numerical Analysis.\\
Hassan II University  of Casablanca, Fstm, Morocco\\
e-mail: khadir@hotmail.com\\
\end{center}

\vspace{0.5cm} \noindent {\bf Abstract:} .\\
In this work, we present a new simple way to encode/decode messages transmitted via a noisy channel and protected against errors by the Hamming method.
We also propose a fast and efficient algorithm for the encoding and  the decoding process which do not use neither the generator matrix
nor the parity-check matrix of the Hamming code.

\vspace{0.2cm} \noindent {\bf Keywords:}  Error-correcting codes, parity-check bits, coding, Hamming code.
\section{Introduction}

The theory of error-correcting codes started in the first half of  the last century with the valuable contribution of Claude Shannon \cite{Shannon1948}.
Richard Hamming \cite{Hamming1950}  was also one of the pioneers in the field. As he pointed it out himself at the end of his article,
the only work on error-correcting codes before his publication, has been that of Golay \cite{Golay1949}.\\
The principle of error-detecting and correcting codes is the redundancy technique. Before sending a message $M$ we code it by adding some bits. Theses bits
are calculated according to some specific mathematical laws. After receiving the sent data, we check again the added bits by the same laws. Therefore,
 we can detect or even correct occurred errors.\\
In 1954, Muller \cite{Muller1954} applied the Boolean algebra for constructing error-detecting codes. In 1960, Reed and Solomon \cite{ReedSolomon1960}
extended the Hamming scheme and proposed  a class of multiple error-correcting codes by exploiting polynomials. Almost at the same year,
Bose with Chaudhuri \cite{BoseChaudhuri1960} and independently Hocquenghem \cite{Hocquenghem1959} showed how to design error-correcting codes if
we choose in advance the number of errors we would like to correct. The method is based on polynomial roots in finite fields.
The invented sets are now known as BCH codes. In 1967, Viterbi  \cite{Viterbi1967} devised a convolutionaly error-correcting code based on the principle
 of the maximum likelihood. In 1978, a cryptosystem based on error-correcting codes theory was proposed by McEliece \cite{McEliece1978}. It is now
  considered  as a serious candidate encryption system that will survive quantum computer attacks \cite{Elder2020}.
Since then, curiously and even with the intensive use of computers and digital data transmission,   no revolutionary change happened in the
methods of error-detecting and correcting codes until the appearance of Turbo-codes in the early 1990s \cite{Berrou1993}.\\
Most of the codes studied in moderne scientific research on coding theory are linear.
They possess a generator matrix for coding and a parity-check matrix for error-correcting. Cyclic codes \cite{Hill1997,Lint1999, Huffman2003}
 constitue a particular remarkable class of linear codes. They are completely determined by a single  binary polynomial and therefore can be
  easily implemented by shift registers.  \\
The low-density parity-check codes, (LDPC), first discovered by
Gallager in 1962 \cite{Gallager1962}, were brought up to date by
MacKay in 1999. These codes have a parity-check matrix whose
columns and rows contain a small number of 1's. Like the Turbo-codes, they achieve
information rates up to the Shannon limit \cite{Berrou1993,MacKay1999,
Shannon1948}.\\
The Hamming code belongs to a family of error-correcting codes whose  coding and decoding procedure are easy. This is why it
is still widely used today in many applications related to the digital data  transmission and commuication networks.
\cite{Lange2001,Otmani2003,Divsalar2004,Stakhov2018,Wang2020,Wu2020,      Huang2021,Li2021}. In 2021, Falcone and Pavone \cite{Falcone2021} studied the weight distribution problem of the binary Hamming code. In 2001, 2015, 2017  and 2018, attempts to improve the decoding of Hamming codes have been performed \cite{Hirst2001,Islam2015,Klein2017,Das2018}.\\
In this work, we present an original and simple way to encode/decode messages transmitted via a noisy channel and protected against errors by the Hamming method.
We consequently construct  algorithms for the encoding and  the decoding procedures which do not use neither the generator matrix nor the parity-check
 matrix of the Hamming code. To the best of our knowledge, this issue
has not been studied before and does not appear in the mathematical or computer science literature on coding theory.\\
The article is organized as follows. In Section 2, we briefly
review the Hamming error-correcting code. Section 3 contains some preliminaries.
Our contribution on coding and decoding messages is described in Section 4. We conclude in Section 5.\\
Throughout this paper, we shall use standard notation. In particular $\mathbb N$ is
the set of all positive integers $0, 1, 2, 3, \dots$. If $a,b,n \in \mathbb N$, we write $a\equiv b\ (mod\ n)$ if $n$ divides $a-b$,
and $a=b\ mod\ n$ if $a$ is the remainder of the Euclidean division of $b$ by $n$. The binary representation of $n$ is
 noted ${\cal B}(n) =\epsilon_{k-1}\epsilon_{k-2}\ldots \epsilon_2\epsilon_1\epsilon_0$ with $\epsilon_i\in\{0,1\}$ and means
 that $\displaystyle n=\sum_{i=0}^{k-1}\epsilon_i 2^i$. The function $\log$  should be interpreted as logarithm to the base 2.
 The largest integer which does not exceed the real $x$ is denoted by $\lfloor x\rfloor$.

\vspace{0.2cm} \noindent
We start, in the next section, by recalling the construction of the Hamming code \cite{Hamming1950} and some known relevant facts on the associated algorithm.
\section{Brief recall on the Hamming code}
Assume that a binary message $M=a_1a_2 \ldots a_n$ was transmitted through a noisy channel and
that the received message is $M'=b_1b_2 \ldots b_n$. If at most one single error has occurred during the transmission, then with $\log n$ parity checks, the
 Hamming algorithm \cite{Hamming1950} efficiently determines the error position or detect double-bit errors. In this section, we review the
 main steps when using the Hamming code in digital data communication networks. There is an abundant literature on the subject. For more
 details, see for instance \cite{Hamming1950}\cite[p.~38]{Lint1999}\cite[p.~319]{Washington2002}\cite[Chap.~8]{Hill1997}\cite[p.~29]{Huffman2003}\cite{Klein2017}
 \cite[p.~23]{MacWilliams1981}.
\subsection{The coding procedure}
In a binary representation, assigning $k$ bits to the error position, allows to analyze and decode any message of length $n=2^k-1$.
The main idea of Hamming method relies on a logical equivalence. The $j$th bit,
 among the $k$ possible bits, is 1 if and only if the error has occurred at a bit $a_i$ of $M$ whose index $i$ has 1 in the $j$th position in
 its binary representation.\\
Hamming defines $a_{2^j}$ as a parity-check bit and it is equal to the sum modulo 2 of all bits of $M$ with index having 1 in the $j$th
position in its binary representation. \\
Bits $a_1,a_2,a_{2^2},\ldots ,a_{2^j},\ldots,a_{2^{k-1}}$ are kept as parity-check bits. All the other bits are for information.
Their number is $m=n-k=2^k-k-1$.
\subsection{The decoding procedure}
Suppose that the  received codeword  is $M'=b_1 b_2 \ldots b_{n}$. For each  fixed integer $j\in \{0,1,2,\ldots,k-1\}$,
if $b_{2^j}$ is the sum modulo 2 of all bits of $M'$ with index having 1 in the $j$th position in its binary representation, then the error position
has 0 at the $j$th  place from the right  in its binary representation, if not, it is a 1.

\vspace{0.2cm} \noindent
{\bf Example 2.1:}
Let us illustrate the technique by un exemple. Suppose that we received a message with 15 bits  as it is indicated in Table 1.

\begin{table}[!ht]
\center
\begin{tabular}{|c|c|c|c|c|c|c|c|c|c|c|c|c|c|c|c|c|}
\hline
Indexes & 1 & 2 & 3 & 4 & 5 & 6 & 7 & 8 & 9 & 10 & 11 & 12 & 13 & 14 & 15 \\
\hline
 bits & 0 & 1 & 1 & 0 & 1 &    0 & 0 & 0 & 1 & 0 &   1 & 1 & 0 & 0 & 1\\
\hline
\end{tabular}
\caption{The received 15 bits}
\end{table}

\vspace{0.3cm}\noindent
Let $\epsilon_3\, \epsilon_2\,  \epsilon_1\,  \epsilon_0$ be the binary representation of the error position. By the Hamming
 algorithm~\cite{Hamming1950}, $\epsilon_0$ is the sum modulo 2 of the bits whose  binary representation starts by 1. So :\\
$\epsilon_0= (b_1+b_3+b_5+b_7+b_9+b_{11}+b_{13}+b_{15})\  mod\  2=1$.\\
$\epsilon_1$ is the sum of the bits whose  binary representation has 1 in the second position. So :\\
$\epsilon_1= (b_2+b_3+b_6+b_7+b_{10}+b_{11}+b_{14}+b_{15})\  mod\  2=0$.\\
$\epsilon_2$ is the sum of the bits whose  binary representation has 1 in the third position. So :\\
$\epsilon_2= (b_4+b_5+b_6+b_7+b_{12}+b_{13}+b_{14}+b_{15})\  mod\  2=1$.\\
$\epsilon_3$ is the sum of the bits whose  binary representation has 1 in the last position. So :\\
$\epsilon_3= (b_8+b_{9}+b_{10}+b_{11}+b_{12}+b_{13}+b_{14}+b_{15})\  mod\  2=0$.\\
Finally, the error position is $\epsilon_3\, \epsilon_2\,  \epsilon_1\,  \epsilon_0=0\  1\  0\  1$ or 5 in the decimal base. The term $b_5$ must be corrected.
\subsection{Complexity of the algorithm }
A Hamming code can also be defined by its $k\times n$ generator
matrix $G$ and its $(n-k)\times n$  parity-check matrix $H$
\cite{Viterbi1979}. Let $x$ be the message to send.
To compute the codeword $xG$, we need $(k-1)n\simeq n\log n$ binary additions and $kn$ bit multiplications.  \\
The decoding complexity : Let $y$ be the received message.
To compute the syndrome $Hy$ we perform $(n-k)(n-1)\simeq n(n-\log n)$ binary additions and $(n-k)n\simeq n(n-\log n)$ bit multiplications.
  To locate the erreur position, in the worst case, we compare the vector $Hy$ to every  column, so  we need  $n-k\simeq n-\log n$ bits comparison.
  As there are $n$
columns in the  matrix $H$, the total number of the bit comparisons is $n(n-k)\simeq n(n-\log n)$. \\
if the columns of parity-check matrix $H$ are arranged in order of
increasing binary numbers from 1 to $n$, we do not need to make
comparisons \cite[p.~83]{Hill1997}. The syndrome $Hy$ is exactly
the binary representation of the error position.


\section{Preliminaries}
Let $k\in \mathbb N-\{0\}$ and $n=2^k-1$.  For every fixed $j\in\{0,1,2, \ldots, k-1\}$ we define the set $S(j,n)$ as
\begin{equation}
S(j,n)=\{0\leq u \leq n \ |  \ {\cal B}(u) \ {\rm contains }\  1\  {\rm in\  the}\ (j+1){{\rm th}}\  {\rm position\  from \ the\  right} \}.
\end{equation}
Hence  $S(0,n)=\{1,3,5,7,\ldots,n  \}$, \\
$S(1,n)=\{2,3,6,7,10,11,\ldots,n  \}$, \\
$S(2,n)=\{4,5,6,7,12,13,14,15,\ldots,n  \}$,\\
$S(3,n)=\{8,9,10,11   ,12,13,14,15,  24,25,26,27,28,29,30,31,       \ldots,n  \}$,\\
$\ldots$\\
Observe that the binary representation of $n$ is ${\cal B}(n)=111\ldots 11$ and so it contains 1 at any position.

\noindent
{\bf Remark 3.1:}
To schematize what is the set $S(j,n)$, suppose that we have in front of us the line representing all positive integers. From the term $2^j$,
we keep the $2^j$ consecutive integers, we delete the following $2^j$ elements, we keep the $2^j$ following, we delete the following $2^j$ terms,
and so on alternately... We repeat this procedure until the part just before the limit $n$.

\vspace{0.2cm} \noindent
{\bf Proposition 3.1:} For any positive integer $u=(2\alpha+1)2^j+i$ where $0\leq  \alpha \leq 2^{k-j-1}-1$ and $0\leq  i \leq 2^j-1$, we have
$u\in S(j,n)$.
\proof First $u$ is a positive integer and $u\leq (2.2^{k-j-1}-2+1)2^j+2^j-1=2^k-1=n$. On the other hand, in base 2, elements $\alpha$ and $i$ can be written as:\\
$\alpha=\displaystyle \sum_{t=0}^m b_t2^t$, $b_t \in \{0,1\}$, $m <k-j-1$ and $i=\displaystyle \sum_{t=0}^r c_t 2^t$, $c_t \in \{0,1\}$, $r <j$. Therefore:\\
$u=\displaystyle \sum_{t=0}^s b_t2^{j+1+t}+2^j+\sum_{t=0}^r c_t 2^t$. As $i\in \{0,1,2,\ldots,2^j-1\}$,  the binary representation of $u$ is
\begin{equation}
{\cal B}(u)=b_s\,b_{s-1}\,\ldots b_1\, b_0\,1\,0\,0 \,  \ldots \,  0\,0\,c_r\,c_{r-1}\,\ldots c_1\, c_0
\end{equation}
As it is easy to see that $b_0$ is in the position $j+2$, we deduce that 1 is in the $(j+1)$th position and the proof is achieved.
\qed

\vspace{0.2cm} \noindent
The next result is essential to the construction of our algorithm.

\vspace{0.2cm} \noindent
{\bf Theorem 3.1:} Let $k\in \mathbb N-\{0\}$ and $n=2^k-1$.  For every fixed $j\in\{0,1,2, \ldots, k-1\}$, if
we set $T(j,n)=\{(2\alpha+1)2^j+i \ | \ 0\leq \alpha \leq 2^{k-j-1}-1 {\rm \ and\ } 0\leq i \leq 2^j-1  \}$, then we have :
\begin{equation}
 T(j,n)=S(j,n)
\end{equation}
\proof Proposition 2.1 shows that $T(j,n) \subset S(j,n)$. Conversely consider an element $u\in S(j,n)$.
So the binary representation  ${\cal B}(u) \ {\rm contains }\  1\  {\rm in\  the}\ (j+1)^{th}\  {\rm position\  from \ the\  right}$. Hence :
$u=\displaystyle \sum_{t=0}^s b_t2^{j+1+t}+2^j+\sum_{t=0}^r c_t 2^t$, $r<j$. By choosing  $\displaystyle \alpha=\sum_{t=0}^s b_t2^{t}$ and
$\displaystyle i=\sum_{t=0}^r c_t 2^t$, we get $u=(2\alpha+1)2^j+i$. Since  $0\leq u\leq n$, we can easily  verify that $0\leq \alpha \leq 2^{k-j-1}-1$,
which ends the proof.
\qed

\vspace{0.3cm} \noindent
{\bf Corollary 3.1:} Let $r=2^j-1$ and $s=2^{k-j-1}-1$. In the  coding Hamming algorithm, the checking bits $a_{2^j}$,
which is artificially zero before the calculation, can be computed as:
\begin{equation}
 \displaystyle a_{2^j}= ( \sum_{\alpha=0}^{s}  \sum_{i=0}^{r}  a_{(2\alpha+1)2^j+i}) \ mod\ 2
\end{equation}
For the decoding step,
\begin{equation}
 \displaystyle \epsilon_{j}= ( \sum_{\alpha=0}^{s}  \sum_{i=0}^{r}  a_{(2\alpha+1)2^j+i}) \ mod\ 2
\end{equation}
is the bit in the $(j+1)$ position from the right of the binary representation of the error location.

\proof By the Hamming algorithm, we have:\\
 $\displaystyle a_{2^j}=(\sum _{u\in S(j,n)-\{2^j\}} a_u)\ mod\ 2=(\sum _{u\in T(j,n)-\{2^j\}} a_u) \ mod\ 2 $, and\\
$\displaystyle \epsilon_{j}=(\sum _{u\in S(j,n)} a_u)\ mod\ 2=(\sum _{u\in T(j,n)} a_u) \ mod\ 2 $,
which give relations (4) and (5).
\qed

\vspace{0.3cm} \noindent {\bf Example 3.1:}  In the early 1980s, the Minitel system \cite[p.~177,185]{Rousseau2008}
\cite[p.~110]{BrunoMartin2004}
was a national network in France,  precursor to the moderne
Internet. A Hamming code  with 7 parity-check bits was used
to correct single error in messages $M=a_1a_2\ldots a_n$, $n=2^7-1$. Let us
see how, by Corollary 3.1,
 we can compute for instance the parity-check bit $a_{2^4}$ in the coding step.\\
We have $n=127$ and $j=4\Longrightarrow   \ r=15 {\rm \  and  \ } s=3$. By relation (4) we fill the following table:
\begin{table}[!ht]
\center
\begin{tabular}{|c|c|}
\hline
 Values of $\alpha$ & bits to add \\
\hline
 0 & $a_{16}=0+a_{17}+a_{18}+\ldots+a_{31}$\\
\hline
1 & $a_{48}+a_{49}+a_{50}+\ldots+a_{63}$\\
\hline
2 & $a_{80}+a_{81}+a_{82}+\ldots+a_{95}$\\
\hline
3 & $a_{112}+a_{113}+a_{114}+\ldots+a_{127}$\\
\hline

\end{tabular}
\caption{Computation of the parity-check $a_{2^4}$}
\end{table}

\noindent
To determine the term $a_{2^4}$, we need to calculate the sum modulo 2 of all the 64 bits in the second column of Table 2.\\
If the message $M$ was received, in the decoding step,  the sum modulo 2 of all the 64 bits in Table 2, with the real received value of $a_{16}$,
gives the $4$th bit from the right  in the binary representation of the erreur position.

\vspace{0.2cm} \noindent
{\bf Theorem 3.2:} Let $k\in \mathbb N-\{0\}$ and $n=2^k-1$.  For every fixed $j\in\{0,1,2, \ldots, k-1\}$, if
we set $U(j,n)=\{2^j+2i-(i \ mod \ 2^j) \ | \ 0\leq  i \leq 2^{k-1}-1  \}$, then we have :
\begin{equation}
 U(j,n)=T(j,n)
\end{equation}
\proof Let $u=2^j+2i-(i \ mod \ 2^j)\in U(2^j,n)$. Put $i=q2^j+r$ with $q\in \mathbb N$   and $0 \leq r <  2^j$. So $i \ mod \ 2^j=r$. Therefore
$u=2^j+q2^{j+1}+2r-r=q2^{j+1}+2^j+r$ and then  ${\cal B}(u)$ has 1 in the $(j+1)$th position from the right. Consequently
$u\in S(j,n)=T(j,n)$ by Theorem 3.1.\\
Conversely let $u=(2\alpha+1)2^j+i\in T(j,n)$. By the definition of $T(j,n)$, we have $0\leq \alpha\leq 2^{k-j-1}-1$ and $0\leq i\leq 2^j-1$.\\
Put $i_1=\alpha 2^j+i$. So $i_1\  mod \ 2^j=i$. On the other hand $2^j+2i_1-(i_1\  mod \ 2^j)=2^j+\alpha 2^{j+1}+2i-i=(2\alpha+1)2^j+i$. Moreover
$0\leq i_1\leq 2^{k-1}-2^j+i\leq 2^{k-1}-2^j+(2^j-1)=2^{k-1}-1$. Conclusion: $u\in U(j,n)$.
\qed

\vspace{0.2cm} \noindent
{\bf Corollary 3.2:} Let $k\in \mathbb N-\{0\}$, $n=2^k-1$ and $r=2^{k-1}-1$.  For every fixed $j\in\{0,1,2, \ldots, k-1\}$, we put $J=2^{j}$.\\
For the coding step:
\begin{equation}
 \displaystyle a_{J}= [ \sum_{i=1}^{r}   a_{J+2i-(i \ mod \ J)}] \ mod\ 2
\end{equation}
For the decoding step:
\begin{equation}
 \displaystyle \epsilon_{J}= [ \sum_{i=0}^{r}   a_{J+2i-(i \ mod \ J)}] \ mod\ 2
\end{equation}
\proof By the Hamming algorithm, for the coding process,  we have
\begin{equation}
\displaystyle a_{2^j}=(\sum _{u\in S(j,n)-\{2^j\}} a_u)\ mod\ 2
\end{equation}
But Theorem 3.1 and Theorem 3.2 imply that the three sets $S(j,n)$, $T(j,n)$, $U(j,n)$
are identical, so:\\
$\displaystyle a_{2^j}=(\sum _{u\in U(j,n)-\{2^j\}} a_u)\ mod\ 2$
$\displaystyle=(\sum _{i=0, i\ne 2^j}^{2^{k-1}-1} a_{2^j+2^i-(i\ mod\ 2^j)})\ mod\ 2$\\
$\displaystyle=(\sum _{i=1}^{2^{k-1}-1} a_{2^j+2^i-(i\ mod\ 2^j)})\ mod\ 2=(\sum _{i=1}^{r} a_{J+2^i-(i\ mod\ J)})\ mod\ 2$,\\
 and we get relation (7).\\
Similar proof for relation (8).
\qed

\vspace{0.2cm} \noindent
The next result is an other  alternative manner for the calculation of the parity-check bits in the Hamming algorithm.\\
{\bf Corollary  3.3:} With the same hypothesis as in Corollary  3.2., we have:\\
For the coding step:
\begin{equation}
 \displaystyle a_{J}= [ \sum_{i=1}^{r}  a_{J(1+\lfloor i/J\rfloor)+i }] \ mod\ 2
\end{equation}
For the decoding step:
\begin{equation}
 \displaystyle \epsilon_{J}= [ \sum_{i=0}^{r}  a_{J(1+\lfloor i/J\rfloor)+i }] \ mod\ 2
\end{equation}
\proof for every $i\in\{0,1,2,\ldots, r\}$, the Euclidean division of $i$ by $J$ gives $i=J\lfloor i/J\rfloor+ (i \ mod \ J)$, so
$J+2i-(i \ mod \ J)=J+i+(\lfloor i/J\rfloor)J+i=J(1+\lfloor i/J\rfloor)+i$. Thus
 $a_{J+2i-(i \ mod \ J)}=a_{J(1+\lfloor i/J\rfloor)+i}$ and by Corollary 3.2, we get relations (10) and (11).
\qed

\vspace{0.2cm} \noindent
We now move to the presentation of our coding and decoding algorithms.

\newpage
\section{Our algorithms for the Hamming code}

Relation (4) in Corollary 3.1 leads to the following coding algorithm where comments are in italic font and
delimited with braces
\begin{algorithm}
\caption{Determination of the parity check symbols }
\begin{algorithmic}
\REQUIRE The message $M=a_1a_2\ldots a_n$ to code before  sending.
\ENSURE The computation of all the checking bits $a_{2^j}$.
\STATE
$k \leftarrow 4$ \{$k$ is the  number of the checking bits
$a_{2^j}$.\} \STATE $n \leftarrow 2^k-1$   {\it \{$n$ is the
length of the message M.\}} \STATE $M \leftarrow [1,1,0,0,0, 0,0,
1,0, 0 , 1, 1, 1, 0, 1]$  {\it \{$M$ is an example of a  message
to code.\}} \FOR{$j \ in \  \{0,1,...,k-1\}$} \STATE $u\leftarrow
2^j$ \{$u$ is the index of the checking symbols $a_{2^j}$.\} \STATE
$max\_alpha \leftarrow 2^{k-j-1}-1$ \{The bound $max\_alpha$ is
the maximal  value of  $\alpha$.\} \STATE $S\leftarrow 0$
\FOR{$\alpha \ in \  \{0,1,...,max\_alpha\}$} \STATE $v \leftarrow
(2\alpha+1)u$ \FOR{$i \ in \ \{0,1,...,2^j-1\}$} \STATE $w
\leftarrow v+i$ \{$w=(2\alpha+1)2^j+i$ is the index of the bit to
add to $S$.\} \STATE $S \leftarrow S+a_w$ \ENDFOR \ENDFOR \STATE
$S \leftarrow S-a_{2^j}$  \{The term $a_{2^j}$ must not be part of
the sum $S$.\} \STATE $S \leftarrow S \mod 2$ \STATE $a_{2^j}
\leftarrow S$  \{We assign $S$ to the checking term $a_{2^j}$.\}
\ENDFOR \STATE print(M) \{$M$ is the final coded message to
send.\}
\end{algorithmic}
\end{algorithm}

\newpage
Relation (5) in Corollary 3.1 leads to the following decoding algorithm where comments are in italic font and
delimited with braces
\begin{algorithm}
\caption{Determination of the error location}
\begin{algorithmic}
\REQUIRE The received message $M=b_1b_2\ldots b_n$ to correct.
\ENSURE The computation of the error position.
\STATE $k \leftarrow 4$ \{$k$ is the  number of the checking bits $b_{2^j}$.\}
\STATE $n \leftarrow 2^k-1$   {\it \{$n$ is the length of the message M.\}}
\STATE $M \leftarrow [1,1,0,0,0, 0,0, 1,0, 0 , 1, 1, 1, 0, 1]$  {\it \{$M$ is an example of a  received message.\}}
\STATE $X \leftarrow 0$ {\it  \{$X$ is the error position in decimal base\}}
 \FOR{$j \ in \  \{0,1,...,k-1\}$}
 \STATE $u\leftarrow
2^j$ \{$u$ is the index of the checking symbols $b_{2^j}$.\}
\STATE $max\_alpha \leftarrow 2^{k-j-1}-1$ \{The bound $max\_alpha$ is
the maximal  value of  $\alpha$.\}
\STATE $S\leftarrow 0$
\FOR{$\alpha \ in \  \{0,1,...,max\_alpha\}$}
\STATE $v \leftarrow (2\alpha+1)u$
\FOR{$i \ in \ \{0,1,...,2^j-1\}$}
\STATE $w\leftarrow v+i$ \{$w=(2\alpha+1)2^j+i$ is the index of the bit to add to $S$.\}
\STATE $S \leftarrow S+b_w$
\ENDFOR
\ENDFOR
 \STATE $S \leftarrow S \mod 2$ \{ $S$ is computed in base 2.\}
  \STATE $X\leftarrow X+S*2^j$  \{We find the error position $X$ in base 10.\}

\ENDFOR \STATE print(X) \{$X$ is the final error position. If $X=0$ there is no error in the transmission.\}

\end{algorithmic}
\end{algorithm}

\section{Conclusion}
In this paper, we presented a new simple and effective method for coding/decoding any transmitted message through a noisy channel that is protected
against errors by the Hamming scheme. We also implemented  practical corresponding algorithms.
Our technique constitues an alternative to the classical use of the  generator matrix for coding or the parity-check matrix for decoding.


\begin{thebibliography}{8}



\bibitem{Berrou1993}
Berrou, C.,  Glavieux, A., and Thitimajshima,  P., {\it Near Shannon limit error-correcting coding and decoding: Turbo-codes},
Proceedings of ICC '93 - IEEE International Conference on Communications, vol. 2, pp1064-1070, (1993).\\
\url{http://citeseerx.ist.psu.edu/viewdoc/download?doi=10.1.1.135.9318&rep=rep1&type=pdf}



\bibitem{BoseChaudhuri1960}
Bose, R.C.,  and Ray-Chaudhri, D.K,  {\it On a class of error-correcting binary group codes},
Inform. Contr., vol. 3, pp68-79, (1960).


\bibitem{Das2018}
Pankaj Kumar Das, {\it Error-locating codes and extended Hamming code},  Matematicki Vesnik, 70,1, pp89-94, (2018).\\
\url{http://elib.mi.sanu.ac.rs/files/journals/mv/271/mvn271p89-94.pdf}

\bibitem{Divsalar2004}
Divsalar, D.,  and Dolinar, S.,  {\it Concatenation of Hamming codes and accumulator
codes with high-order modulations for high-speed decoding}, IPN Progress Report 42-156, (2004).\\
\url{https://www.researchgate.net/publication/245759463_Concatenation_of_Hamming_Codes_and_Accumulator_Codes_with_High-Order_Modulations_for_High-Speed_Decoding}

\bibitem{Elder2020}
Elder, J.,
{\it Quantum resistant Reed Muller codes on McEliece cryptosystem} Thesis, Phd, University of North Carolina, USA, (2020).\\
\url{https://math.charlotte.edu/sites/math.charlotte.edu/files/fields/preprint_archive/paper/2020_01.pdf}

\bibitem{Falcone2021}
Falcone, G., and Pavone, M., {\it Binary Hamming codes and Boolean designs}, Designs, Codes and Cryptography, 89, pp1261-1277, (2021).\\
\url{https://www.researchgate.net/publication/350768990_Binary_Hamming_codes_and_Boolean_designs}


\bibitem{Gallager1962}
Gallager, R.G., {\it Low-density parity-check codes}, IRE Trans. Inf.
Theory, vol. 8, no. 1, pp. 21-28, (1962).\\

\bibitem{Golay1949}
Golay, M.J.E.,  {\it Notes on digital coding}, Proceedingsof the I.R.E.,
Vol. 37, pp657, (1949).\\
\url{http://www.lama.univ-savoie.fr/pagesmembres/hyvernat/Enseignement/1617/info528/TP-Golay/golay_paper.pdf}


\bibitem{Hamming1950}
Hamming, R., {\it Error-detecting and error-correcting codes}, Bell Syst. Tech. J. 29, pp147-160, (1950).

\bibitem{Hill1997}
Hill, R., {\it A first course in coding theory}, Oxford University Press, (1997).


\bibitem{Hirst2001}
Hirst, S.,  Honary, B., {\it A simple soft-input/soft-output decoder for Hamming codes}, Cryptography and coding,
pp38-43, Lecture Notes in Comput. Sci., 2260, Springer, Berlin, (2001).

\bibitem{Hocquenghem1959}
Hocquenghem, A., {\it Codes correcteurs d'erreurs}, Chiffres,
Vol. 2, pp147-156, (1959).


\bibitem{Huffman2003}
Huffman, W.C., and Pless, V., {\it Fundamentals of error-correcting codes}, Cambridge University Press, (2003).

\bibitem{Huang2021}
Jianhong Huang, Guangjun Xie, Rui Kuang, Feifei Deng, Yongqiang Zhang, {\it QCA-based Hamming code circuit for nano communication network},
Microprocessors and Microsystems, 84, pp1-12,  (2021).


\bibitem{Islam2015}
Islam, M.S., Kim, C.H., and Kim, J.M.,  {\it Computationally efficient implementation of a Hamming code decoder using graphics processing unit},
Journal of Communications and Networks. Institute of Electrical and Electronics Engineers (IEEE), (2015).\\
\url{https://www.researchgate.net/publication/269935106_Computationally_Efficient_Implementation_of_a_Hamming_Code_Decoder_Using_Graphics_Processing_Unit}

\bibitem{Klein2017}
Klein, S.T., Shapira, D.,  {\it Hierarchical parallel evaluation of a Hamming code}, Algorithms (Basel), 10, (2017).

\bibitem{Lange2001}
Lange, C., and Ahrens, A., {\it On the undetected error probability for shortened Hamming codes on channels with memory}, Cryptography and coding, p9-19,
Lecture Notes in Comput. Sci., 2260, Springer, Berlin, (2001).

\bibitem{Li2021}
Li, Lin; Chang, Chin-Chen; Lin, and Chia-Chen {\it Reversible data hiding in encrypted image based on $(7,4)$ Hamming code
and unit smooth detection}, Entropy 23, no. 7, (2021). \\
\url{https://www.ncbi.nlm.nih.gov/pmc/articles/PMC8306628/}

\bibitem{Lint1999}
Lint, van, J.H., {\it Introduction to coding theory}, Third edition, Springer, (1999).

\bibitem{MacWilliams1981}
   MacWilliams, F.J., and  Sloane, N.J.A,  {\it The theory of error-correcting codes}, North-Holland publishing company,
   Third printing, (1981).

\bibitem{BrunoMartin2004}
Martin, B., {\it Codage, cryptologie et applications}, Presses polytechniques universitaire romandes, (2004).

\bibitem{MacKay1999}
MacKay, D.J.C.,  {\it Good error-correcting codes based on very sparse matrices}, IEEE Trans.
Inform. Theory, vol. 45, no. 2, pp399-431, (1999).


\bibitem{McEliece1978}
McEliece, R.J., {\it A public-key cryptosysem based on algebraic coding theory}, DSN Progress Report,  pp42-44,  (1978).


\bibitem{Muller1954}
Muller, D.E., {\it Application of the Boolean algebra to switching circuit design and to error detection},
IRE, Trasaction electronic computers, pp6-12, (1954).

\bibitem{Otmani2003}
Otmani, A.,  {\it Caractérisation des codes auto-duaux binaires de type II
à partir du code de Hamming étendu [8, 4, 4]}, C. R. Acad. Sci. Paris, Ser. I 336 (2003).\\
\url{https://www.sciencedirect.com/journal/comptes-rendus-mathematique/vol/336/issue/12}

\bibitem{ReedSolomon1960}
Reed, I.S., and Solomon, G.,  {\it Polyomial codes over certain finite fields}, J. Soc. Indust. Al. Math.
Vol. 8, No. 2,   pp300-304, (1960).\\
\url{https://faculty.math.illinois.edu/~duursma/CT/RS-1960.pdf}


\bibitem{Rousseau2008}
Rousseau, C., and  Saint-Aubin, Y., {\it Mathematics and Technology}, Springer,  (2008).


\bibitem{Shannon1948}
Shannon, C., {\it A Mathematical theory of communication}, The Bell System Technical Journal,
 Vol. 27, pp379-423 and 623-656,(1948).\\
\url{https://people.math.harvard.edu/~ctm/home/text/others/shannon/entropy/entropy.pdf}


\bibitem{Stakhov2018}
Stakhov, A., {\it Mission-critical systems, paradox of Hamming code, row hammer effect, `Trojan horse' of the binary system and numeral systems with
irrational bases}, Comput. J. 61,  no. 7, (2018).\\
\url{https://academic.oup.com/comjnl/article/61/7/1038/4430323?login=true}

\bibitem{Washington2002}
Trappe, W., and Washington, L.C.,  {\it Introduction to cryptography and coding theory}, Printice Hall, (2002).

\bibitem{Viterbi1967}
Viterbi, A.J., {\it Error bounds for convolutional codes and an asymptotically optimum decoding algorithm},
IEEE Trans. Inf. Theory IT-13, pp260-269 (1967).


\bibitem{Viterbi1979}
Viterbi, A.J.,  Omura, J.K.,  {\it Principles of digital communication and coding}, McGraw-Hill, Inc.,  (1979).


\bibitem{Wang2020}
Yanting Wang, Mingwei Tang, Zhen Wang, {\it High-capacity adaptive steganography based on LSB and Hamming
code}, Inter. J. for Light and Electron Optics, 213, pp1-9,  (2020).

\bibitem{Wu2020}
Xiaotian Wu, Ching-Nung Yang, Yen-Wei Liu, {\it A general framework for partial reversible data hiding using Hamming code},
Signal Processing, 175, pp1-12,  (2020).



\end{thebibliography}
\end{document}